\documentstyle[twoside,fleqn,espcrc2,epsfig]{article}

\newcommand{\ee}{\end{equation}}
\newcommand{\be}{\begin{equation}}
\newcommand{\bea}{\begin{eqnarray}} 
\newcommand{\eea}{\end{eqnarray}}

\newcommand{\plaq}{\Box}

\title{Exact Local Bosonic Algorithm for Dynamical Quarks}
\author{A.Galli$^a$ and Ph. de Forcrand$^b$\\[0.1cm]
{\em $^a$ Max-Planck-Institut f\"ur Physik, D-80805 Munich, Germany}\\
{\em $^b$ SCSC-ETH-Zentrum, CH-8092 Z\"urich, Switzerland}}
\begin{document}
\begin{abstract}
We present an exact local bosonic algorithm for the simulation of
dynamical fermions in lattice QCD. We show that this algorithm is 
a viable alternative to the Hybrid Monte Carlo algorithm.
\end{abstract}

\maketitle
\section{Introduction}

The local bosonic algorithm was proposed by L\"uscher \cite{luescher1} in a hermitian 
version as a new method to simulate full QCD.
Here we present several significant improvements of 
this algorithm:  an inexpensive stochastic Metropolis accept/reject test
\cite{forcrand1,ultimo} to make the algorithm exact;  a non-hermitian polynomial approximation \cite{forcrand1,ultimo} and
 a simple even-odd preconditioning \cite{beat1,ultimo}. 
We illustrate the 
effectiveness of our algorithm for two-flavor QCD and compare its performance with the Hybrid Monte Carlo (HMC) algorithm in the relevant regime of small quark mass. We 
successfully model its static properties (the Monte Carlo acceptance), and try 
to disentangle its dynamics.  
From our analysis it appears that this version of the local bosonic algorithm 
is indeed a viable alternative to the HMC.

\section{Description of the algorithm}

The full QCD partition function with two fermion flavors is given by
\be 
Z=\int [dU] |\det D|^2 e^{-S_G[U]}
\label{fullQCD}
\ee 
where $D$ represents the fermion matrix and $S_G$ the gauge action. This partition function can be approximated by introducing a polynomial
approximation of the inverse of the fermion matrix.  \be |\det
D|^2=\det D^\dagger\det D\simeq \frac{1}{|\det P(D)|^2}
\label{app}
\ee The polynomial $P(z)=\prod_{k=1}^n(z-z_k)$ of degree $n$ is
defined in the complex plane and approximates the inverse of $z$. 
Since we are investigating full QCD with two flavors, the determinant 
$|\det P(D)|^2$ manifestly factorizes into positive pairs, so that the
$\frac{1}{|\det P(D)|^2}$ 
term of the approximation (\ref{app}) can be expressed
as a Gaussian integral over a set of boson fields $\phi_k$ ($k=1,...,n$)
with color and Dirac indices. The full
QCD partition function (\ref{fullQCD}) is then approximated by 
\be
Z\simeq \int [dU][d\phi][d\phi^\dagger]\ e^{-S_L[U,\phi]}
\label{lus}
\ee 
where $\phi$ represents the set of all boson field families,
$S_L$ is the local action $S_L=S_G+S_b$ and 
$
S_b=\sum_{k=1}^n 
|(D-z_k)\phi_k|^2
$.
Making use of the locality of $S_L$ 
we may now simulate the partition function (\ref{lus}) by locally updating
the boson fields and the gauge fields, using heat-bath and
over-relaxation algorithms. \\ 
The simulation of full QCD can be obtained from (\ref{lus}) by correcting
the errors due to the approximation through a Metropolis test at the end of each trajectory.
Introducing the error term in the partition function we obtain 
\be 
Z=\int [dU] [d\phi] [d\phi^\dagger]|\det(DP(D))|^2e^{-S_L(U,\phi)}\label{exact}
\ee

The correction term $ |\det(DP(D))|^2$ can be evaluated by
expressing the determinant $ |\det(DP(D))|^2$
as a Gaussian integral over an auxiliary field $\eta$ and incorporating it in the partition sum (\ref{exact})
\be
Z=\int [dU][d\eta][d\eta^\dagger]
 [d\phi] [d\phi^\dagger]e^{-S_{exact}(U,\phi,\eta)}
\label{exact2}
\ee
by defining a new exact action
\be
S_{exact}(U,\phi,\eta)=S_{L}(U,\phi)+S_{C}(U,\eta)
\ee
where 
$
S_{C}(U,\eta)=|[DP(D)]^{-1}\eta|^2
$
is the correction action. In carrying out the simulation one generates 
configurations of $(U,\phi)$ and $\eta$ such that the probability of 
finding a particular configuration is proportional to $\exp(-S_{exact})$. 
The strategy is to alternatively update the $(U,\phi)$ 
fields and the $\eta$ field. The Metropolis acceptance  probability 
$P_{acc}$ is defined using the exact action $S_{exact}$ 
so that the transition probability of the algorithm 
satisfies detailed balance with respect to $S_{exact}$ \cite{ultimo}. To be more precise:
\bea
&&P_{acc}((U,\phi)\rightarrow(U',\phi'))=\label{df}\\
&&\left\{\begin{array}{ll}
\min\left(1,e^{-|W\chi|^2+|\chi|^2}\right)&\mbox{{\footnotesize if $S_G(U)\geq S_G(U')$}}\\
\min\left(1,e^{+|W^{-1}\chi|^2-|\chi|^2}\right)&\mbox{{\footnotesize if $S_G(U)<S_G(U')$}}
\end{array}\nonumber
\right.
\eea
where $\chi$ is a Gaussian distributed spinor (used to generate $\eta$ with a global heat-bath by $\eta=DP(D)\chi$ or $\eta=D'P(D')\chi$) and $W=[D'P(D')]^{-1}DP(D)$. Here $D$ and $D'$ denote the fermion matrix with the $U$ and $U'$ gauge configurations, 

respectively. 
The algorithm can be summarized as follows:\\[0.1cm]
$\bullet$ Generate a new Gaussian spinor $\chi$ with variance one;\\
$\bullet$ Update locally the boson and gauge fields $m$ times (in a reversible and optimized order \cite{beat1})
according to the approximate partition function (\ref{lus});\\
$\bullet$ Accept/reject the new configuration according to the Metropolis acceptance probability (\ref{df}).\\

In order to evaluate the Metropolis acceptance probability we have to solve 
a linear system involving $DP(D)$ or $D'P(D')$, 
for which we use the BiCGstab algorithm
 \cite{schilling}. This linear system is very well conditioned because $P(D')$ 
(or $P(D)$) approximates the inverse of $D'$ (or $D$). 
The cost for solving it is minimal
and scales like the local updating algorithms in the volume and quark mass.\\

\section{Numerical results}

Numerical simulations using the exact local bosonic algorithm have been performed  for different lattice parameters.
The majority of the simulations are reported in \cite{ultimo}.
We explored the acceptance of the Metropolis correction test and the number of 
iterations of the BiCGStab algorithm used in that test for inverting $DP(D)$.  
The study was performed by varying the degree $n$ of the polynomial and the hopping parameter $k$. 
One observes that the acceptance increases quite rapidly with the degree of 
the polynomial, above some threshold \cite{ultimo} (see Fig.1). On the other hand, the number of 
iterations needed to invert $DP(D)$ remains very low for high enough acceptance. The data show clearly that the overhead due to the Metropolis test remains 
negligible provided that the degree of the polynomial is tuned to have 
sufficient acceptance. 
Results obtained from simulations using 
even-odd preconditioning confirm, as expected, that the improvement of the 
approximation reduces the required number of bosonic fields by at least a factor two \cite{ultimo,beat1}. 
This improvement is trivially implemented by a simple redefinition of the roots $z_k$ \cite{ultimo}.\\
\begin{figure}
\epsfig{file=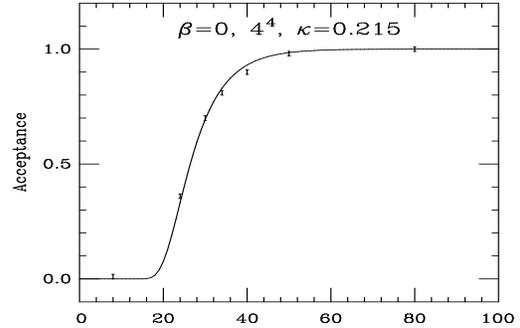,bbllx=5pt,bblly=5pt,bburx=10pt,bbury=12pt,width=7cm,height=7cm,clip=}
\vspace{-2.7cm}
\caption{{\small Acceptance as a function of the number of fields. The line represents the fit ansatz (8).}}
\vspace{-0.8cm}
\end{figure}
We have also performed a simulation for parameters relevant for physical interesting conditions. On a $8^4$ lattice at $\beta=5.6$ and $k=0.1585$ (very near $k_c$) with Schr\"odinger functional boundary conditions we have compared the performance of our a

lgorithm with HMC. In Table 1 we present the plaquette and its integrated autocorrelation time in units of fermion-matrix multiplications ($[D\phi]$) The simulation with the local bosonic algorithm was performed using 
48 boson fields. Its average acceptance is $<P_{acc}>=0.589(1)$.
The HMC data is obtained from \cite{kj}.These results show that for light quarks the peformance of our algorithm is comparable to the performance of HMC.\\[0.1cm]
{\small
\begin{tabular}{|c|l|l|}\hline
 & $<\plaq>$ & $\tau_{int}(\plaq)$\\\hline
HMC & 0.5726(1) & 33750(6000) $[D\phi]$\\
multiboson & 0.57275(15) & 23040(4000) $[D\phi]$ \\\hline
\end{tabular}
Table 1
}

\section{Optimisation}

Using some general assumptions and the known error bounds on the Chebyshev-like approximation of $D^{-1}$, we can obtain \cite{ultimo} an ansatz
for the acceptance probability, which takes the form
\begin{equation}
< P_{acc} > = erfc\left( f \sqrt{96 V} 
\left(\frac{k}{k_c}\right)^{n + 1} \right)
\label{fitacc}
\end{equation}
where $V$ is the lattice volume, $k$ is the hopping parameter of the Wilson fermion matrix $D$ and $k_c$ is the critical hopping parameter at a given $\beta$.
We expect the fitting parameter
$f$ to depend smoothly on $\beta$, but very little on $n, V$ and $k$. In \cite{ultimo} we show that our ansatz (\ref{fitacc}) is very satisfactory (see Fig.1).
At other values of $\beta$, one can fix $f$ by a test on a small lattice,
and then predict the acceptance for larger volumes and different quark masses.
We can use our ansatz to analyze and optimize the cost of a simulation. From the analysis it turns out that the optimal average acceptance is around 0.7-0.8.

\section{Scaling}

The total complexity with
the volume and the quark mass is related to the scaling of the number of bosonic fields and the autocorrelation of the observables.
As the volume $V$ increases, the number $n$ of bosonic fields should
grow like $\log V$. 
As the quark mass $m_q$ decreases, the number of fields $n$ must grow like
$m_q^{-1}$. This is a consequence of the exponential convergence of the
polynomial approximation. We expect the autocorrelation to behave like $\tau\sim nm_q^{-\alpha}$ \cite{beat1,ultimo,borrelli} so that the complexity is proportional to $n^2m_q^{-\alpha}$. The exponent $\alpha$ can only be derived from MC data since
there is no clear theoretical understanding of the coupled dynamics of the gauge and boson fields.
 From an exploratory simulation \cite{ultimo} on a $4^4$ 
lattice at $\beta=0$ we obtained 
that $\alpha$ is near $1$. This is only 
indicative, because the dependence of $\alpha$ on the volume and the 
coupling has yet to be explored. 

\section{Conclusion}

We have presented an alternative algorithm to HMC for simulating dynamical 
quarks in lattice QCD. This algorithm is based on a local bosonic action. 
A non-hermitian polynomial approximation of the inverse of the quark matrix 
is used to define the local bosonic action. The addition of a global 
Metropolis test corrects the systematic errors. The overhead of the correction 
test is minimal. This algorithm is exact for any choice of the polynomial approximation. 
No critical tuning of the approximation parameters is needed. Only the 
efficiency of the algorithm, which can be monitored, will be affected by 
the choice of parameters.
The cost of the algorithm scales similarly with that of HMC 
with the volume $V$ of the lattice 
and with the inverse of the quark mass. 
Finally, for light quarks the peformance of our algorithm is comparable to the performance of HMC provided that there is enough memory to store the needed boson fields.

\end{document}